\documentstyle[12pt,epsf]{article}
\newcommand{\beq}{\begin{equation}}
\newcommand{\eeq}{\end{equation}}
\newcommand{\beqa}{\begin{eqnarray}}
\newcommand{\eeqa}{\end{eqnarray}}

\begin{document}

\pagestyle{empty}

\begin{center}

\hfill TK 95 27

\smallskip

{\large { \bf STRANGE TWISTS IN NEUTRAL PION PHOTO/ELECTRO--PRODUCTION}}

\vspace{1.cm}

Ulf--G. Mei{\ss}ner\\
{\it Universit\"at Bonn, ITKP, Nussallee
14-16, D--53115 Bonn, Germany}

\vspace{0.4cm}

\end{center}

\baselineskip 10pt

\noindent I review the interesting tale of the electric dipole amplitude in
neutral pion photoproduction and the resulting consequences. I also discuss why
there is new life related to P--wave multipoles. Electroproduction is
briefly touched upon.

\vspace{0.5cm}

\baselineskip 14pt

\noindent {\bf 1 $\quad$ THE EARLY YEARS}

\vspace{0.3cm}

\noindent Some 25 years ago, de Baenst and
Vainsthein and Zakharov (VZ) [1,2] independently  derived a
so--called low--energy theorem (LET) for the electric dipole amplitude
$E_{0+}$ measured in threshold $\pi^0$ photoproduction off protons,
\beq
E_{0+}(s_{\rm thr}) = -\frac{e g_{\pi N}}{8 \pi m} \, \mu \, \biggl\{
1 - \frac{1}{2} (3 + \kappa_p ) \mu + {\cal O}(\mu^2) \biggr\}
= -2.3 \times 10^{-3} / M_{\pi^+} \, \, \, ,
\eeq
with $s_{\rm thr} = (m+M_\pi)^2$, $\mu \equiv M_\pi /m \simeq 1/7$ and $M_\pi$
($m$) the pion (nucleon) mass as well as $\kappa_p$ the anomalous magnetic
moment of the proton and $g_{\pi N}$ the strong pion--nucleon coupling
constant. The expansion of $E_{0+}(s_{\rm thr})$ in powers of the small
parameter $\mu $ as given in Eq.(1)
will from now on be called the 'low--energy guess' (LEG) [3].
It is of particular
interest since in the chiral limit of vanishing pion mass, $E_{0+}(s_{\rm
thr})$ is zero and thus appears to be a good candidate to test our
understanding of the explicit chiral symmetry breaking in QCD related to the
current quark masses $m_{u,d}$ appearing in the QCD Hamiltonian,
\beq
M_\pi^2 = -(m_u + m_d) <0|\bar{q} q |0> / F_\pi^2 + {\cal O}(m^2_{u,d})
\quad ,
\eeq
with $F_\pi = 93$~MeV the pion decay constant and the scalar quark condensate
is believed to be the order parameter of the spontaneous chiral symmetry
breaking in QCD. The derivation of the LEG supposedly only assumes
 very general principles like gauge
invariance and PCAC. That, however, is not quite correct. VZ [2] stressed
 that an
extra analyticity assumption has been made. They even checked the validity of
this by calculating the rescattering diagram  and found it to hold true for
what was believed to be the largest correction to the Born terms leading to
Eq.(1). The current commutator algebra manipulations used by de Baenst [1]
were effectively hiding this subtlety.
For a long time, the LEG was dormant since the existing data on
threshold neutral pion production off protons were not very accurate but
reassuringly close, $E_{0+}^{\rm exp} (s_{\rm thr})= -1.8 \pm 0.6$
(in natural units which I will drop from now on). So one had to wait some time
for a serious test of the LEG.

\vspace{0.5cm}

\noindent {\bf 2 $\quad$ SHOCK, RELIEVE AND THE SPOILERS}

\vspace{0.3cm}

\noindent The papers of the Saclay [4] and Mainz [5] groups
both claimed a substantial
deviation from the LEG by many standard deviations. As usual, theorists were
(too) quick to invent ways to modify the LEG or claiming the discrepancy to be
a measure of the light quark masses. To avoid embarrasment, I will not give
references here. It was also pointed out, by some experimenters and theorists,
that there were some flaws in the interpretation of the data. In case of the
Saclay results, the large rescattering contribution had been incorrectly
subtracted. Reinstating that, one finds $-1.5 \pm 0.3$, where the error is
a guess. In the Mainz case, the ambiguity in the two solutions could be
resolved by imposing the constraint of the total cross section. That results in
$E_{0+}(s_{\rm thr}) = -2.0 \pm 0.2$ (see e.g. [6,7]), in satisfactory
agreement with the LEG prediction, Eq.(1). Paradise seemed to be regained.

However, there was a problem. In 1991, V\'eronique Bernard, J\"urg Gasser,
Norbert Kaiser and I published a paper in which we showed that based on chiral
perturbation theory (CHPT), which is the effective field theory of the Standard
Model (SM) at low energies,
the expression Eq.(1) is modified at order $\mu^2$. The
correct low--energy theorem (LET) reads [8]
\beq
E_{0+}(s_{\rm thr}) = -\frac{e g_{\pi N}}{8 \pi m} \, \mu \, \biggl\{
1 - \biggl[ \frac{1}{2} (3 + \kappa_p )+ \bigl(\frac{m}{4F_\pi}\bigr)^2
\biggr] \mu + {\cal O}(\mu^2) \biggr\}
\, \, \, ,
\eeq
The physics underlying this new term at next--to--leading order is well
explained in Ref.[8], it simply amounts to a breakdown of the analyticity
assumption made by VZ [2], see also the discussion in Refs.[9,10]. Without an
explicit loop calculation, this effect at order $\mu^2$ could not have been
found. What is distracting, however, is the fact that the coefficent
of the second term is now so large that it cancels the leading one and thus
even leads to a positive value for $E_{0+}(s_{\rm thr})$. Consequently, the
form of the LET as given in Eq.(3) can not be used to test the chiral dynamics
of QCD. Also, what has clouded the discussion for a long time was the
accidental closeness of the reexamined Mainz data with the LEG prediction,
Eq.(1).

\vspace{0.5cm}

\noindent {\bf 3 $\quad$ CONVERGENCE AT LAST ?}

\vspace{0.3cm}

\noindent CHPT has tought us that reactions involving S--wave interactions
between pions or pions and nucleons often require higher order calculations
to remove discrepancies between theory and experiment, like in the scalar
form factor of the pion, the reaction $\gamma \gamma \to \pi^0 \pi^0$,
$K \to \pi^0 \gamma \gamma$ and alike. It can therefore be expected that
the lowest order one loop calculation leading to Eq.(3) is not sufficiently
accurate. Thus, Bernard, Kaiser and I performed a higher order calculation
for the electric dipole amplitude, i.e. the pertinent S--wave multipole [11].
At that order ($p^4$ in the chiral counting, where $p$ denotes a small
momentum), one has to consider one loop graphs with insertions from the
dimension two effective pion--nucleon Lagrangian and counter terms of the
type
\beq
{\cal L}^{\rm ct} = e \, a_1 \, \omega \, M_\pi + e \, a_2 \, M_\pi^3
\, \, \, ,
\eeq
with $\omega$ the energy of the pion in the $\pi N$ cms. In the threshold
region, the pion three--momentum is very small, $\vec{q \,}_\pi \simeq 0$ and
thus $\omega \simeq M_\pi$. Naturalness of the low--energy constants $a_{1,2}$
lets us assume that $a_1 \simeq a_2 = {\cal O}(1)$ so that effectively only the
sum $a_1 + a_2$ counts in the threshold region. Fitting the Mainz data and
 letting
$a_1$ and $a_2$ completely free, one finds $a_1 \simeq -a_2 \simeq
50$~GeV$^{-4}$ with $a_1 + a_2 = 6.7$~GeV$^{-4}$ one order of magnitude
smaller. On the other hand, if one restricts the values of these low--energy
constants by resonance exchange (in this case $\Delta, \, \rho^0$ and $\omega$
excitation),  one finds more natural numbers $a_1 \simeq a_2
\simeq 3$~GeV$^{-4}$ with almost the same sum as in the free fit [11]. So there
is a clear discrepancy which might  be due to (a) some higher order effects
or (b) some inconsistency in the data or (c) a combination of both. Resonance
exchange saturation of the low--energy constants, which is well established in
the meson sector, leads us to believe that $E_{0+}(s_{\rm thr}) \simeq -1.2$
(with some large uncertainty which is hard to quantify in the absence of a two
loop calculation). To resolve this puzzle, the experimenters come to our
rescue. The new Mainz data of Fuchs et al. [12], shown here by Thomas Walcher
in his talk, exhibit a clear reduction of the total cross section below the
$\pi^+ n$ threshold and show a good agreement with the CHPT calculation based
on resonance exchange, see Fig.1. The new experimental value, also corroborated
by the SAL measurements [13], knocks the LEG off by many standard deviations,
\beq
E_{0+}^{\exp}(s_{\rm thr}) = -1.33 \pm 0.08 \quad .
\eeq
While one might consider the agreement with the CHPT prediction of $-1.2$
accidental, the energy-dependence in the threshold region fits also with the
CHPT result based on resonance exchange. In particular, the small value of
$E_{0+}$ at $\pi^+ n$ threshold ($\simeq -0.4$) is a clear indication of chiral
loops. A simple Born model with form factors can never explain such a trend.
It is gratyfying to finally have an experimental verification of the expected
reduction of the electric dipole amplitude due to loop effects. Good news is
that the LEG is out, the bad one is that the original hope of quantitatively
testing chiral dynamics in the S--wave has been chattered. But that's not yet
the end of the story, fortunately.

\hskip 1.9in
\epsfysize=3in
\epsffile{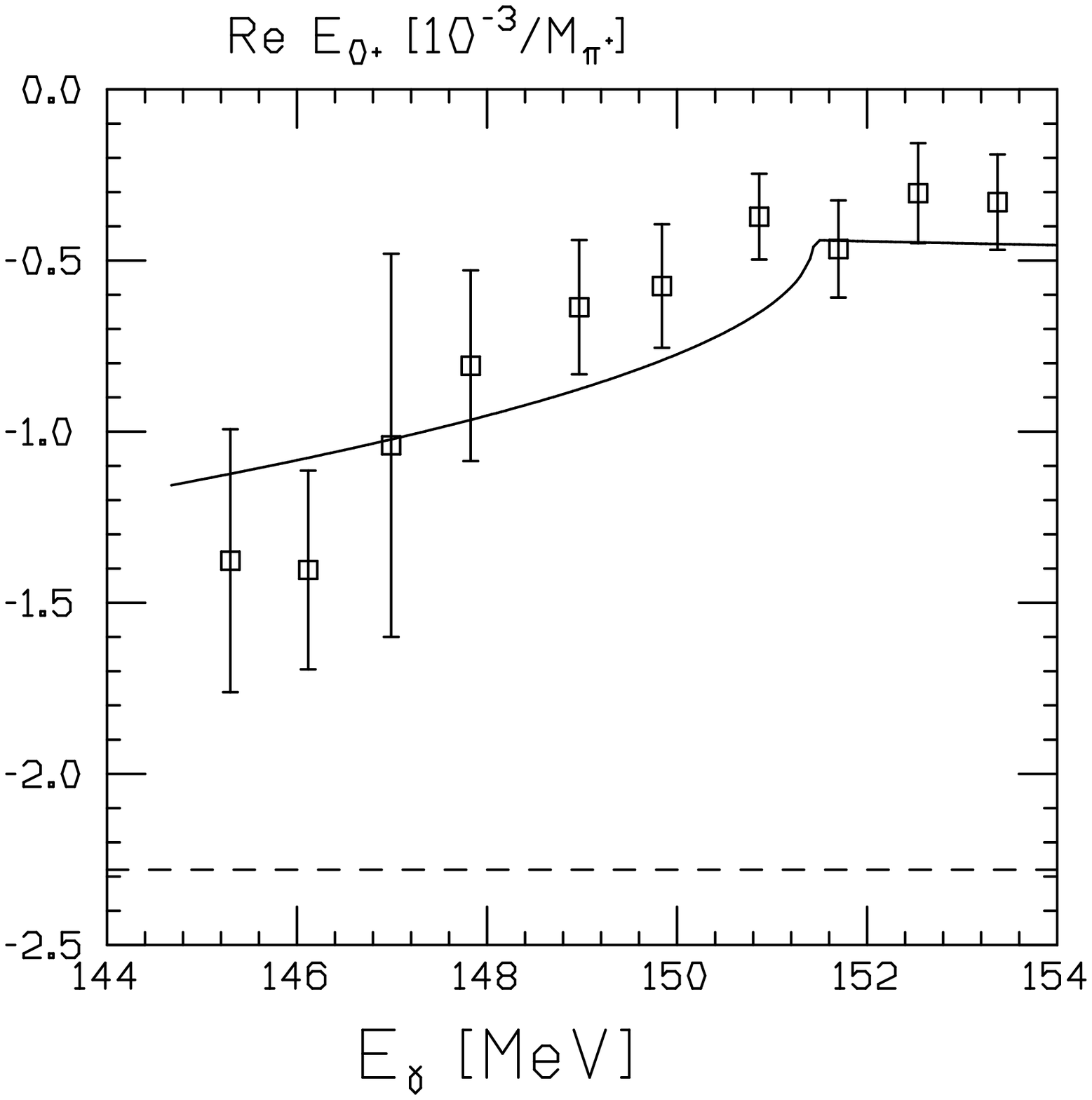}


\centerline{Fig.~1:\quad  Re~$E_{0+}$: CHPT prediction [11] versus data [12].}


\medskip 

\vspace{0.5cm}

\noindent {\bf 4 $\quad$ P--WAVES ARE INTERESTING ? YES, YES \& YES !}

\vspace{0.3cm}

\noindent A quick look in the textbooks shows the P--wave dominance of the
total cross section very quickly after threshold. This is, of course, the
excitation of the $\Delta(1232)$. Therefore, it is generally believed that
chiral dynamics is of no relevance for the three P--waves, $M_{1+}$,
$M_{1-}$ and $E_{1+}$ ($E/M$ for electric/magnetic, $l=1$ for P--wave and
$\pm$ for $j = l \pm 1/2$ the total angular momentum of the final $\pi N$
system). However, in the transition matrix element the combinations
\beq
P_1 = 3 E_{1+} + M_{1+} - M_{1-} \, , \, \,
P_2 = 3 E_{1+} - M_{1+} + M_{1-} \, , \, \,
P_3 = 2 M_{1+} + M_{1-} \, \, \, ,
\eeq
appear naturally. One quickly realizes that $P_3$ is indeed completely
dominated by $\Delta$ and vector meson contributions.
Not so for $P_1$ and $P_2$
-- a back on the envelope calculation shows that $\Delta$--exchange drops out
to leading order (using the static $\Delta$ well--known from the $\Delta$--hole
model). I strongly encourage the sceptical reader to perform this little
exercise. It is also worth to stress that already 20+x years ago, Balachandran
and collaborators [14]
noted this and made predictions for the slopes of $P_{1,2}$ at
threshold. However, the method   used in [14]
only gave the leading term and was only
applicable to the isoscalar amplitudes (thier method, out--dated by now, could
neither give  the isovector amplitude nor any next--to--leading
order correction). In Ref.[11] novel LETs
were derived for the slopes of $P_1$ and $P_2$. Consider first $P_1$,
\beq
{1 \over |\vec q\,|} P_{1,{\rm thr}}^{\pi^0 p}
 = {e g_{\pi N} \over 8 \pi m^2}
\biggl\lbrace 1 + \kappa_p + \mu \biggl[ -1 - {\kappa_p \over 2} + {g_{\pi N}^2
(10 - 3\pi ) \over 48 \pi} \biggr] + {\cal O}(\mu^2) \biggr\rbrace \, \, ,
\eeq
and similarly for $P_2$
\beq
{1 \over |\vec q\,|} P_{2,{\rm thr}}^{\pi^0 p}
 = {e g_{\pi N} \over 8 \pi m^2}
\biggl\lbrace -1 - \kappa_p + {\mu \over 2} \biggl[ 3 + \kappa_p
 - {g_{\pi N}^2  \over 12 \pi} \biggr]
+ {\cal O}(\mu^2) \biggr\rbrace \, \, .
\eeq
We note that the P--waves scale with the pion momentum, and not with the
product of the pion times the photon momentum as commonly assumed, see also
Ref.[14]. Eqs.(8,9) are examples of  quickly converging $\mu$ expansions,
\beq
{1 \over |\vec q\,|} P_{1,{\rm thr}}^{\pi^0 p}  = 0.512 \, ( 1 - 0.062 )
 \, {\rm GeV}^{-2} = 0.480 \, {\rm GeV}^{-2} \, \, , \eeq
and
\beq
{1 \over |\vec q\,|} P_{2,{\rm thr}}^{\pi^0 p}  = -0.512 \, ( 1 - 0.0008 )
 \, {\rm GeV}^{-2} = -0.512 \, {\rm GeV}^{-2} \, \, . \eeq
Similar expressions for the neutron can be found in Ref.[11].
Only $P_1$ can be inferred in a model--independent manner from the unpolarized
data. The new Mainz analysis [12] leads to
\beq
{1 \over |\vec q\,|} P_{1,{\rm thr}}^{\pi^0 p}
  = 0.47 \pm 0.01 \, {\rm GeV}^{-2} \, \, ,
\eeq
in stunning agreement with the LET prediction. One can also combine the
predictions for $P_1$ and $P_2$ to disentangle the magnetic from the electric
piece. That has been done by Jack Bergstrom [15] using  data from coherent
neutral pion photoproduction of $^{12}$C together with the old Mainz data
for $\gamma p \to \pi^0 p$. He
finds a good agreement for the magnetic LET $\sim M_{1+} - M_{1-}$ but a
sizeable deviation for the electric one ($\sim E_{1+}$). Note that for this
latter quantity the large leading term $\sim (1+\kappa_p)$ cancels out and one
is considering the small difference of two small numbers. A direct measurement
of the photon asymmetry $\Sigma(\theta)$
underway at MAMI will help to settle this issue.

\vspace{0.5cm}

\noindent {\bf 5 $\quad$ SOME REMARKS ON THE NEUTRON}

\vspace{0.3cm}

\noindent In Ref.[11] the electric dipole amplitude for the reaction
$\gamma n \to \pi^0 n$ was also calculated, using the low--energy constants
determined from resonance exchange. The updated version for this quantity
based on the new data from Mainz [12] to fix the counter term coefficients
is shown in Fig.2. The stunning result is that it is quite a bit larger in
magnitude than the corresponding proton one. This result seems
counterintuitive if one uses the classical dipole argument to estimate the
relative strength of the electric dipole amplitude in charged and

\hskip 2in
\epsfysize=3in
\epsffile{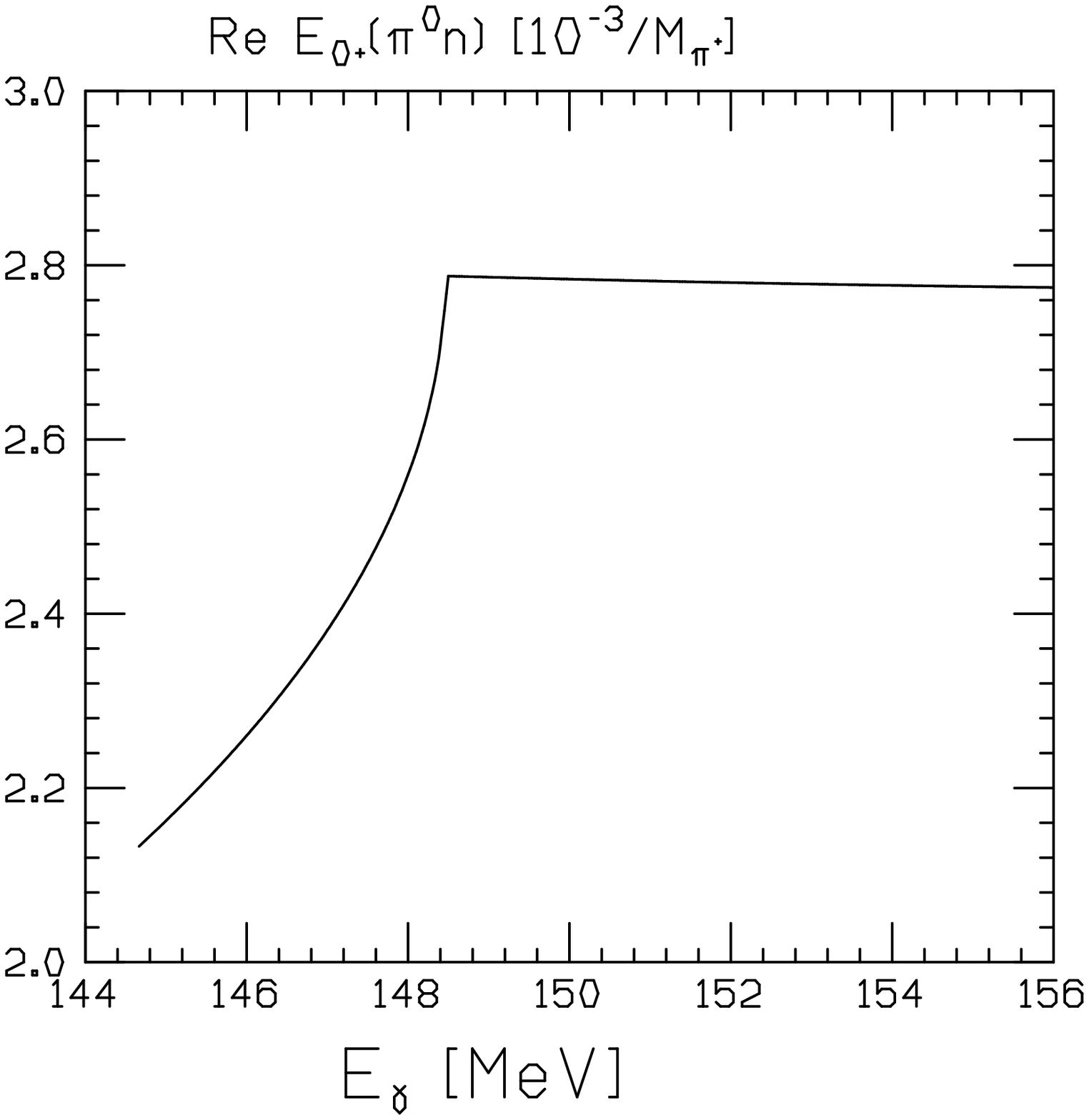}


\vspace{-0.5cm}

\centerline{Fig.~2:\quad  CHPT prediction for Re$E_{0+}$ in $\gamma n \to
                          \pi^0 n$.}

\medskip 

\noindent   neutral
pion production. Quantum physics, however, is not always adhering to such
notions as exemplified here. It would therefore be very important to measure
this quantity. In the absence of pure neutron targets, this is a difficult
job. For example, in the deuteron the large charge--exchange amplitude has
to be accounted for very precisely before one could get at the elementary
$n \pi^0$ amplitude [16], compare also the   CHPT calculation for $\pi^0$
production off the deuteron by Beane et al. [17].
This is certainly a third generation experiment.
However, it would be very important to have independent information on all
photoproduction channels, $\gamma p \to \pi^+ n$, $\gamma n \to \pi^- p$,
$\gamma p \to \pi^0 p$ and $\gamma n \to \pi^0 n$, since that would ultimately
lead to a test of isospin symmetry. Needless to say, a more systematic
treatment
of the pion--nucleon--photon system including virtual photon loops and the
quark
mass difference $m_u - m_d$ has to be done before such data could be
interpreted correctly. This is a challenge to both the experimenters and the
theorists.

\vspace{0.5cm}

\noindent {\bf 6 $\quad$ NEUTRAL PION ELECTROPRODUCTION}

\vspace{0.3cm}

\noindent A rapidly developing field is the extension to electroproduction.
This is motivated by the facts that (a) virtual photons can also couple
longitudinally to the nucleon spin and (b) that all multipoles depend on the
energy and the momentum transfer. Therefore, much richer information can be
obtained  in comparison to the case of real photons.
First measurements at $k^2 = -0.1$~GeV$^2$ (where $k^2$ is the photon
virtuality) performed at NIKHEF were reported in Refs.[18,19]. The S--wave
cross section extracted in [18] can be understood within the framework of
relativistic CHPT [20]. However, in that case one has no consistent power
counting and is thus limited in accuracy. In particular, the S--wave multipoles
$E_{0+}$ and $L_{0+}$ have to be calculated to higher orders. Then, two
new counter terms appear (at order $p^4$) which can e.g. be fixed from a best
fit to the differential cross section data of Ref.[19]. The preliminary
result of such a fit is shown in Fig.3. For the resonance fit,
i.e. estimating these two constants from $\Delta$ and vector meson
excitation, there is one
new $N\Delta \gamma$ coupling parametrized by a coupling constant $g_3$
and an off--shell parameter $X'$ [21]. One notices that the free
(denoted by the dash--dotted line in Fig.3) and
the resonance fit (solid line)  are roughly consistent (but not exactly).
We remark that in Ref.[19] the
electroproduction P--wave LETs presented in Ref.[22]
(these are the generalizations of Eqs.(7,8) for virtual

\hskip 1in
\epsfysize=5in
\epsffile{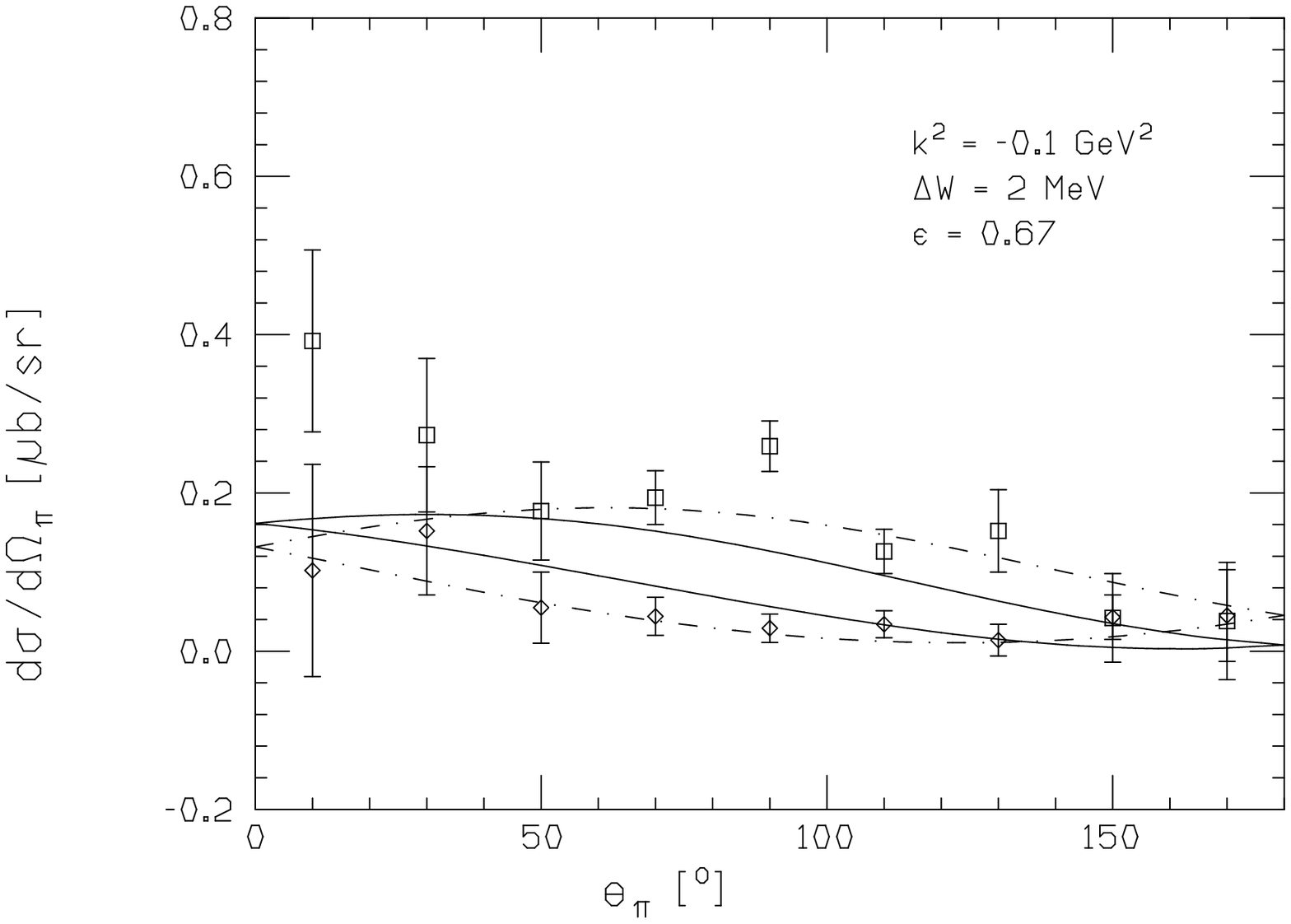}

\baselineskip 10pt

\vspace{-1.5cm}

\centerline{Fig.~3:\quad  CHPT fit to the data of Ref.[19], see text.}

\medskip 

\baselineskip 14pt

\noindent  photons) were used to
extract the S--wave multipoles. Our preliminary analysis seems to lead to
somewhat smaller values for $E_{0+} (\Delta W)$ (with $\Delta W$ the energy
above the pion--nucleon threshold), somewhere between $+2$ and $+ 2.5$ for
$\Delta W = 2 \ldots 15$ MeV. It is worth to stress that the electric
dipole amplitude has changed sign as compared to the photoproduction
case. The longitudinal multipole $L_{0+}$ stays negative from the
photon point up to $k^2 = -0.1$ GeV$^2$.  More precise data have been taken at
MAMI and were shown by Thomas Walcher here [23]. The preliminary analysis of
these data (also at $k^2 = -0.1$~GeV$^2$) seems to indicate an
L/T ratio quite
consistent with Born terms but not with the relativistic CHPT result. However,
what one really should compare to are the improved $p^4$ calculations which
will soon be available (with the new low--energy constants fixed from
the NIKHEF data, cf. Fig.3). Clearly, the real test of the CHPT
prediction will be the comparison to the MAMI data taken at lower $k^2$.
These data have
not yet been analyzed. Of particular interest is the value of $k^2$ where
the electric dipole amplitude changes sign. There is much more to come in terms
of quantitative comparisons between theory and experiment.

\vspace{0.5cm}

\noindent {\bf 7 $\quad$ SUMMARY AND OUTLOOK}

\vspace{0.3cm}

\noindent For the case of neutral pion photo/electroproduction,
let me summarize the recent developments as follows:

\begin{itemize}

\item[$\bullet$] In the S--wave, the effect of chiral loops clearly shows up
on a {\it qualitative} level. For a precise test of chiral dynmaics, one would
need to perform much more accurate calculations presumably by a synopsis with
dispersion theory. This remains to be done.

\smallskip

\item[$\bullet$] The real {\it quantitative} tests of chiral dynamics are
related to the {\it P-waves}. This is an important new result which comes quite
unexpectedly. For the large P--wave multipoles, rather accurate calculations
exist and improvement is needed for the small multipoles like $E_{1+}$. On the
experimental side, polarization experiments will help to disentangle the
small from the large multipoles due to much increased sensitivities. Such
experiments are either planned or underway.

\smallskip

\item[$\bullet$] In $\pi^0$ electroproduction, accurate data are just becoming
available and the same holds true for a more refined theoretical description.
In the very near future, there will be a huge body of data to be compared with
theoretical expectations. In particular, measurements at smaller photon
virtuality, say at $k^2 = -0.05$~GeV$^2$, are urgently called for.

\smallskip

\end{itemize}

Other topics of interest I could not address in  detail are the photo-- and
electroprodcution of charged pions, to find out the deviations from the
leading Kroll--Ruderman LET and to pin down the axial form factor $G_A (k^2)$
at low momentum transfer. Furthermore, precise kaon and eta production data
have been taken and partly been published. The extension to the
three--flavor sector
is not trivial due to the (a) closeness of some resonances in certain channels
and (b) the more sizeable symmetry breaking effects due to the larger
$K$ and $\eta$
masses. With respect to the second problem, some progress has been made
recently in a complete $p^4$ calculation of the baryon octet masses and the
pion--nucleon $\sigma$--term, see Ref.[24]. Furthermore, there now seems to
be a consistent method of implementing the $\Delta$ in the effective field
theory as discussed by Joachim Kambor [25]. This method can then be used to
further extend the range of applicability of CHPT and to calculate e.g. the
much discussed $E2/M1$ ratio measured at the resonance position.

\bigskip

Finally, to appreciate the rapid progress made in this field I recommend to
read the summary which Berthold Schoch and I wrote in the summer of last year
[26] -- it is quite amazing to see how much theory and experiment have improved
and the resulting  shift of emphasis is also noteworthy.

\vspace{1.5cm}

\noindent {\bf 8 $\quad$ ACKNOWLEDGEMENTS}

\vspace{0.3cm}

\noindent I would like to thank V\'eronique Bernard and Norbert Kaiser for
making the BKM collaboration so fruitful and allowing me to present some
preliminary results. They can, however, not be made responsible for any
error in my presentation. I am also grateful to Jack Bergstrom and Michael
Fuchs for informing me about their data prior to publication.

\newpage

\noindent{\bf References}

\bigskip

\noindent 1. P. de Baenst, {\it Nucl. Phys.} {\bf B24}, 633 (1970).

\smallskip

\noindent 2. A.I. Vainsthein and V.I. Zakharov,
{\it Nucl. Phys.} {\bf B36}, 589 (1970).

\smallskip

\noindent 3. G. Ecker, in "Chiral Dynamics: Theory and Experiments",
A.M. Bernstein and

B.R. Holstein (eds), Springer, Heidelberg, 1995.

G. Ecker and Ulf--G. Mei{\ss}ner,
{\it Comm. Nucl. Part. Phys.} {\bf 21} (1995) 347.

\smallskip

\noindent 4. E. Mazzucato {\it et al.},
 {\it Phys. Rev. Lett.\/} {\bf 57}, 3144 (1986).

\smallskip

\noindent 5. R. Beck {\it et al.}, {\it Phys. Rev. Lett.\/}
{\bf 65}, 1841 (1990).

\smallskip

\noindent 6. A. M. Bernstein and B. R. Holstein,
{\it Comm. Nucl. Part. Phys.\/} {\bf 20}, 197 (1991).

\smallskip

\noindent 7. J. Bergstrom, {\it Phys. Rev.\/} {\bf C44}, 1768 (1991).

\smallskip

\noindent 8. V. Bernard, J. Gasser, N. Kaiser and Ulf-G. Mei{\ss}ner,
{\it Phys. Lett.\/} {\bf B268}, 219 (1991).

\smallskip

\noindent 9. V. Bernard, N. Kaiser, and Ulf-G. Mei{\ss}ner, {\it Nucl. Phys.\/}
{\bf B383}, 442 (1992).

\smallskip

\noindent 10. V. Bernard, N. Kaiser, and Ulf-G. Mei{\ss}ner,
{\it Int. J. Mod. Phys.\/} {\bf E4}, 193 (1995).

\smallskip

\noindent 11. V. Bernard, N. Kaiser and Ulf-G. Mei{\ss}ner, {\it Z. Phys.}
{\bf C}  (1995) in press.

\smallskip

\noindent 12. M. Fuchs et al., Gie{\ss}en preprint, submitted to
{\it Phys. Lett.} {\bf B} (1995).

\smallskip

\noindent 13. J. Bergstrom, private communication.

\smallskip

\noindent 14. A.P. Balachandran et al.,  {\it Ann. Phys.\/} (NY)
{\bf 45}, 339 (1967).

\smallskip

\noindent 15. J. Bergstrom, {\it Phys. Rev.\/} {\bf C52}, xxx (1995).

\smallskip

\noindent 16. G. Miller, private communication.

\smallskip

\noindent 17. S. Beane, C.Y. Lee and U. van Kolck,  {\it Phys. Rev.\/}
{\bf C} (1995) in print.

\smallskip
\noindent 18. T. P. Welch et al., {\it Phys. Rev. Lett.\/}
{\bf 69}, 2761 (1992).

\smallskip

\noindent 19. H.B. van den Brink et al.,
Phys. Rev. Lett. {\bf 74}, 3561 (1995).

\smallskip

\noindent 20. V. Bernard, N. Kaiser, T.-S. H. Lee and Ulf-G. Mei{\ss}ner,

{\it Phys. Rev. Lett.} {\bf 70}, 387 (1993).

\smallskip

\noindent 21. V. Pascalutsa and O. Scholten, {\it Nucl. Phys.\/}
{\bf A591}, 658 (1995).

\smallskip

\noindent 22. V. Bernard, N. Kaiser and Ulf-G. Mei{\ss}ner,
{\it Phys. Rev. Lett.} {\bf 74}, 3752 (1995).

\smallskip

\noindent 23. M. Distler et al., in preparation; Th. Walcher, this conference.

\smallskip

\noindent 24. B. Borasoy and Ulf-G. Mei{\ss}ner, {\it  Phys. Lett.}
{\bf B} (1995) in press.

\smallskip

\noindent 25. J. Kambor, this conference.

\smallskip

\noindent 26. Ulf--G. Mei{\ss}ner and B. Schoch, in "Chiral Dynamics: Theory
and Experiments",

A.M. Bernstein and  B.R. Holstein (eds), Springer,
Heidelberg, 1995.

\end{document}